# Diagnostic Uncertainty Limits the Potential of Early Warning Signals to Identify Epidemic Emergence

Callum R.K. Arnold[1,2,*], Matthew J. Ferrari[1,2]

[1] Department of Biology, Pennsylvania State University, University Park, PA, USA 16802
[2] Center for Infectious Disease Dynamics, Pennsylvania State University, University Park, PA, USA 16802

---

[*] Corresponding author. Callum R.K. Arnold. Address: Department of Biology, Pennsylvania State University, University Park, PA, USA 16802. Email: contact@callumarnold.com.

# Abstract

Methods to detect the emergence of infectious diseases, and approach to the "critical transition" $R_{\mathrm{E}} = 1$, have to potential to avert substantial disease burden by facilitating preemptive actions like vaccination campaigns. Early warning signals (EWS), summary statistics of infection case time series, show promise in providing such advanced warnings. As EWS are computed on test positive case data, the accuracy of this underlying data is integral to their predictive ability, but will vary with changes in the diagnostic test accuracy and the incidence of the target disease relative to clinically-compatible background noise. We simulated emergent and null time series as the sum of an SEIR-generated measles time series, and background noise generated by either independent draws from a Poisson distribution, or an SEIR simulation with rubella-like parameters. We demonstrate that proactive outbreak detection with EWS metrics is resilient to decreasing diagnostic accuracy, so long as background infections remain proportionally low. Under situations with large, episodic, noise, imperfect diagnostic tests cannot appropriately discriminate between emergent and null periods. Not all EWS metrics performed equally; we find that the mean was the least affected by changes to the noise structure and magnitude, given a moderately accurate diagnostic test ($\geq 95\%$ sensitive and specific), and the autocovariance and variance were the most predictive when the noise incidence did not exhibit large temporal variations. In these situations, diagnostic test accuracy should not be a precursor to the implementation of an EWS metric-based alert system.

**Key words:** Diagnostic Test Uncertainty; Infectious Disease Surveillance; Outbreak Detection; Early Warning Signals; Critical Slowing Down.

# Background

Despite sustained advances over decades, infectious diseases still pose a substantial threat to human life, causing an estimated 33.8 B infections, and 57.0 M deaths, per annum in 2019 (rising to an estimated 67.9 M deaths in 2021, as a result of the COVID-19 pandemic) [1]. For many diseases, effective and affordable vaccines have played a substantial role in reducing this burden, averting 154 million deaths since the introduction of the Expanded Programme on Immunization in 1974 [2]. As burden decreases with increasing control, dynamics may shift from predictable annual incidence to increasingly variable and episodic dynamics [3]. Many populations that have achieved apparent control suffer from large-scale resurgent outbreaks due to the build up of susceptibles in the absence of persistent transmission [4–6]. While rapid detection and response has the potential to minimize the impact of these outbreaks [7–10], early warning systems that can trigger preemptive action prior to outbreaks would have a greater potential for effectiveness.

Infectious disease surveillance systems are crucial for detecting outbreaks [11,12], and could be leveraged to anticipate the risk of outbreaks [13–15]. Outbreak detection and response systems are reactive in nature; cases are collated, counted, and if a pre-determined threshold is met or breached, an action is undertaken (e.g., preliminary investigation, or reactive vaccination campaign) [16,17]. However, due to the exponential trajectory of incidence in the early stages of an outbreak, the reactive nature necessarily results in excess infections that cannot be prevented [7–9]. To limit the burden of disease, ideally, epidemiologists could utilize the output of a surveillance system (e.g., the trend in cases of a pathogen) to predict the risk of a future outbreak, triggering a *proactive* action, such as a preventative vaccination campaign.

The risk of an outbreak can be quantified in terms of the effective reproduction number, $R_{\mathrm{E}}$, defined as the expected number of secondary cases due to each infectious individual [18]. $R_{\mathrm{E}} = 1$ represents a "critical transition", below which epidemics should not spread, and above which outbreaks should propagate. There has been growing interest, in many fields, to identify and develop early warning signals (EWS) that are predictive of the approach to such critical transitions in dynamical systems [19–23]. The appeal of an alert system based upon EWS metrics is that they are model-free, only requiring the calculation of summary statistics of a time series. Prior work has demonstrated that for infectious disease systems, computing EWS metrics on the progression of population susceptibility may be most predictive [15], but collecting this information is often intractable, and utilizing either the incidence or prevalence data has provided similarly useful predictions [13,15,24]. If an EWS is predictive, critical slowing down theory suggests that the EWS values will change as a transition is approached, such as an increase in the variance. Prior work has demonstrated that EWS metrics are theoretically correlated with a critical transition for infectious disease systems, under emergent and extinction conditions [13–15,24–26].

While identifying EWS that are correlated with a transition is an important first step, systems to preempt outbreaks also require a discrete decision threshold to trigger preventive action (e.g., vaccination) [27,28]. To address this, various threshold-based and statistical learning based approaches have been developed [22,29–32]. For these, a distribution of the EWS metric is quantified during a non-outbreak regime and a decision threshold is triggered when the EWS metrics at time $t$ exceeds some quantile or derived statistic of this distribution; often the mean plus 2 times the standard deviation. Prior work has shown that a single exceedance is often too sensitive and requiring multiple consecutive flags to trigger an alert improves the accuracy in a 'noisy' system by reducing the false positive rate [29,33,34].

Until now, the relatively nascent topic of EWS for outbreak detection has only explored imperfect surveillance in the setting of under-reporting and temporal aggregation of case data [13,35]. Our goal is to characterize the performance of EWS metrics for outbreak detection in a surveillance system with diagnostic uncertainty due to co-circulating pathogens and imperfect diagnostic tests, i.e., non-target disease that may be misdiagnosed as the target disease. For diseases with non-specific symptoms, e.g., measles and rubella that often co-circulate and have similar clinical presentation [17,36], an imperfect diagnostic test will result in false positive and negative cases. In this paper we show the conditions under which diagnostic uncertainty overwhelms the time series used to calculate EWS summary statistics, limiting the ability to predict epidemic transitions.

## Materials & Methods

### Model Structure

We modeled the dynamics of a target pathogen (measles), for which we want to detect outbreaks, with a stochastic compartmental non-age structured Susceptible-Exposed-Infected-Recovered (SEIR) model. The SEIR model was simulated using a Tau-leaping algorithm with a time step of 1 day, with binomial draws so that no jump resulted in negative compartment sizes [37,38]. We assumed no seasonality in the transmission rate ($\beta_t$), and set the latent and infectious periods equal to 10 days and 8 days, respectively, and an $R_0$ equal to 16, approximating measles parameters values [39,40]. Demographic parameters (birth and death rates) broadly reflecting those observed in Ghana were selected to evaluate the performance of EWS metrics in a setting where high, yet sub-elimination, vaccination coverage is observed, requiring ongoing vigilance [36,41]. An initial population of 500,000 individuals was simulated, with commuter-style imports drawn from a Poisson distribution with mean proportional to the size of the population and $R_0$, to maintain a level of endemicity [42].

To evaluate the predictive ability of EWS metrics in environments with background disease that could produce false positive test results if tested with an imperfect diagnostic, we generated a time series of "suspected measles" by summing the measles and background noise time series. The noise time series is modeled as either: independent draws of a Poisson distribution, with mean equal to a multiple (c) of the daily average measles incidence, where $c \in \{1, 7\}$; or from an SEIR time series with rubella-like parameters with additional noise drawn from a Poisson distribution with mean equal to 15% of the daily average of the rubella incidence time series, to account for non-rubella sources of clinically-compatible febrile rash e.g., parvovirus (Table 1) [43,44]. Under dynamical (SEIR-generated) noise simulations, the vaccination rate at birth was selected to produce equivalent magnitudes of daily average noise incidence as observed in the static noise simulations that were drawn from Poisson distributions (10.20% and 87.34%). Throughout the rest of the manuscript, these will be referred to as low and high dynamical/static noise scenarios, accordingly. Each day, all clinically-compatible febrile rash cases (that is, both the measles and noise time series) were tested using one of the following diagnostic tests, producing a time series of test positive cases.

- A perfect test with 100% sensitivity and specificity. This was chosen to reflect the best-case scenario that the imperfect diagnostic-based alert scenarios could be compared against.
- An imperfect diagnostic with sensitivity and specificity equal to either 99%, 98%, 97%, 96%, 95%, 90%, or 80%.

Table 1: Compartmental model parameters

| Parameters | Measles - Emergent | Measles - Null | Dynamical noise |
|---|---|---|---|
| R0 | 16 | | 5 |
| Latent period (s) | 10 days | | 7 days |
| Infectious period (g) | 8 days | | 14 days |
| Vaccination rate at birth during burn-in period ($r_i$) | Unif (92.69%, 100%) | | 10.20%, 83.74% |
| Vaccination rate at birth after burn-in period ($r_e$) | Unif (60%, 80%) | Unif (92.69%, 100%) | 10.20%, 83.74% |
| Birth/death rate (m) | 27 per 1000 per annum | | |
| Importation rate | $\frac{1.06 * \mu * R_0}{\sqrt{N}}$ | | |
| Population size (N) | 500,000 | | |
| Initial proportion susceptible | 0.05 | | |
| Initial proportion exposed | 0.0 | | |
| Initial proportion infected | 0.0 | | |
| Initial proportion recovered | 0.95 | | |

To evaluate the performance of the EWS metrics at predicting the approach to the critical transition ($R_E = 1$) from below, we simulated "emergent" scenarios where $R_E$ increases until 1, and "null" scenarios where $R_E$ is below 1. For both emergent and null scenarios, we generated 100 time series. All measles simulation incorporated a 5-year burn-in period to produce sufficient data for calculation of the EWS metrics upon aggregation, as well as to produce greater variation in the trajectory of $R_E$. For each time series, the vaccination rate at birth during the burn-in period was sampled from a Uniform distribution between 92.69% and 100% coverage. These bounds were selected to ensure the maximum value of $R_E$ that could be reached within 10 years (twice the length of the burn-in period) was 0.9. We simulated emergent scenarios by lowering the vaccination rate at birth after completion of the burn-in period, allowing the proportion of the population that is susceptible to grow. For each emergent time series, the vaccination rate at birth was independently drawn from a Uniform distribution between 60% and 80% coverage, allowing the rate of growth in $R_E$, and therefore the time of the critical transition, to vary in each emergent time series. For each null time series, the vaccination rate at birth was set to the coverage sampled during the burn-in period, ensuring $R_E$ would not cross the critical transition within the scope of the simulation, though it may grow slowly. Each of the 100 emergent and 100 null time series are paired during the pre-processing steps i.e., up until the completion of the burn-in period, paired emergent and null simulations share the same vaccination rate at birth, and they are both truncated to identical lengths (the time step when $R_E = 1$ in that pair's emergent simulation).

All simulations and analysis was completed in Julia version 1.10.5 [45], with all code stored at https://github.com/arnold-c/CSDNoise.

### Computing & Evaluating EWS

Each set of null and emergent time series are aggregated by month and numerical estimates of the EWS metrics were then calculated on the aggregated time series, detrended using backwards-facing moving averages with bandwidth $b = 52$ weeks. For example, the EWS metric, the mean, is given by the expectation in Equation 1 where: $X_s$ represents the aggregated incidence at time point (month) $s$, and $\delta = 1$ time step (in the simulation results presented, 1 month). At the beginning of the time series when $t < b$, $b$ is set equal to $t$.

$$\hat{\mu}_t = \sum_{s=t-b\delta}^{s=t} \frac{X_s}{b} \qquad\qquad 1$$

In this paper we evaluate the performance of the following EWS metrics: the mean, variance, coefficient of variation, index of dispersion, skewness, kurtosis, autocovariance, and autocorrelation at lag-1, which have previously been shown to be correlated or predictive of disease emergence [13,15,27,27,35]. The full list of numerical formulas for each EWS metric can be found in Table 2.

Table 2: Numerical computations for EWS metrics, where $\delta = 1$ time step, $b = 52$ weeks

| EWS Metric | Formula |
|---|---|
| Mean ($\hat{\mu}_t$) | $\sum_{s=t-b\delta}^{s=t} \frac{X_s}{b}$ |
| Variance ($\hat{\sigma}_t^2$) | $\sum_{s=t-b\delta}^{s=t} \frac{(X_s - \hat{\mu}_s)^2}{b}$ |
| Coefficient of Variation ($\widehat{\text{CV}}_t$) | $\frac{\hat{\sigma}_t}{\hat{\mu}_t}$ |
| Index of Dispersion ($\widehat{\text{IoD}}_t$) | $\frac{\hat{\sigma}_t^2}{\hat{\mu}_t}$ |
| Skewness ($\widehat{\text{Skew}}_t$) | $\frac{1}{\hat{\sigma}_t^3} \sum_{s=t-b\delta}^{s=t} \frac{(X_s - \hat{\mu}_s)^3}{b}$ |
| Kurtosis ($\widehat{\text{Kurt}}_t$) | $\frac{1}{\hat{\sigma}_t^4} \sum_{s=t-b\delta}^{s=t} \frac{(X_s - \hat{\mu}_s)^4}{b}$ |
| Autocovariance ($\widehat{\text{ACov}}_t$) | $\sum_{s=t-b\delta}^{s=t} \frac{(X_s - \hat{\mu}_s)(X_{s-\delta} - \hat{\mu}_{s-\delta})}{b}$ |
| Autocorrelation lag-1 ($\widehat{\text{AC-1}}_t$) | $\frac{\widehat{\text{ACov}}_t}{\hat{\sigma}_t \hat{\sigma}_{t-\delta}}$ |

Once the EWS metrics have been computed, the correlation within emergent time series is computed using Kendall's Tau-B, signifying if an EWS metric consistently increases (or decreases) in magnitude throughout the time series [46,47]. Kendall's Tau is computed on two lengths of time series: from the beginning of the simulation until the critical transition is met, and from the completion of the burn-in period until the critical transition is met. To evaluate the strength of the correlation, we use the area under the receiver operator curve (AUC), as described in prior papers [13,24,27]. Briefly, the calculation of the AUC compares whether the distributions of Kendall's Tau differ substantially between emergent and null simulations for a given alert scenario and EWS metric. AUC is calculated using the rank order of the EWS metrics for both emergent and null time series using Equation 2 [48] where $r_{\text{null}}$ equals the sum of ranks for the null time series, and $n_{\text{null}}$ and $n_{\text{emergent}}$ refer to the number of null and emergent simulations, respectively. An AUC of 0.5 indicates the EWS is similarly correlated with both emergent and null time series, offering no benefit; values > 0.5 indicate a positive correlation with emergent time series, and < 0.5 indicates the EWS metric is negatively correlated with the emergent simulations. AUC values are transformed as $|\text{AUC} - 0.5|$ to highlight the strength of the correlation with emergence, with values close to 0 exhibiting poor performance, and a value of 0.5 indicating perfect correlation [13].

$$\text{AUC} = \frac{r_{\text{null}} - n_{\text{null}}(n_{\text{null}} + 1)/2}{n_{\text{emergent}} n_{\text{null}}} \qquad 2$$

The primary mode of evaluation for the EWS metrics relies on computing and triggering an alert based upon a set of alert conditions. The alert scenario is defined as the combination of diagnostic test, noise structure and magnitude, and EWS metric. The combination of the alert scenario and EWS alert hyperparameters (the quantile threshold, Q, value of the long-running metric distribution that must be exceeded to create a flag, and the number of consecutive flags, C, required to trigger an alert), produce distinct counts of emergent and null time series that result in an alert. For example, a simulation may require that at two consecutive time points (C = 2), the corresponding values of the EWS are larger than 95% of previously observed EWS values (Q = 0.95). The sensitivity of the system is defined as the proportion of the emergent simulations that result in an alert, and the specificity is the proportion of the null simulations that do not result in an alert. Taking the mean of the sensitivity and specificity produces the accuracy of the system. For each alert scenario, a grid search over the EWS hyperparameters ($Q \in [0.5, 0.99]$, $C \in [2, 30]$) is performed to identify the set of EWS hyperparameters that maximizes alert accuracy for a given alert scenario. If multiple hyperparameter combinations produce identical alert system accuracies, the combination with the highest alert specificity is selected. After the optimal EWS hyperparameters have been selected, the accuracy of each EWS metric are compared across alert scenarios, at their respective maximal values. Finally, the speed and timing of detection relative to the critical transition is evaluated using Kaplan-Meier survival estimates [49]. Alert accuracy was only evaluated on EWS calculated after the completion of the burn-in period.

## Results

### Correlation with Emergence

The strength and direction of the raw correlation (Tau) between EWS metrics and the approach to the critical transition in emergent time series is strongly dependent upon the length of the time series evaluated; Tau is higher when calculated after the burn-in period for the top 5 ranked metrics (Table 3). Normalizing the correlation in the emergent time series against the correlation observed in null simulations yields comparable results when calculated from the full time series and only after the burn-in (Table 3). Consistent with previous studies, the autocovariance, variance, mean, and index of dispersion show the strongest correlations with emergence ($|\text{AUC} - 0.5| = 0.20, 0.20, 0.18, 0.13$, evaluated after the burn-in period, respectively) [13,35].

Table 3: The ranking and mean value of Kendall's Tau computed on emergent time series, and the $|AUC - 0.5|$ for each metric. The values are computed on the full time series, and the subset from after the completion of the burn-in period, with a perfect test

| Rank | Tau | | \|AUC - 0.5\| | |
|---|---|---|---|---|
| | Full Time Series | After Burn-In Period | Full Time Series | After Burn-In Period |
| 1 | Index of dispersion (0.26) | Variance (0.62) | Mean (0.19) | Autocovariance (0.20) |
| 2 | Autocovariance (0.25) | Index of dispersion (0.58) | Autocovariance (0.17) | Variance (0.20) |
| 3 | Variance (0.25) | Autocovariance (0.58) | Variance (0.17) | Mean (0.18) |
| 4 | Mean (0.22) | Autocorrelation (0.38) | Index of dispersion (0.14) | Index of dispersion (0.13) |
| 5 | Autocorrelation (0.17) | Mean (0.38) | Coefficient of variation (0.12) | Autocorrelation (0.12) |
| 6 | Coefficient of variation (-0.01) | Coefficient of variation (0.15) | Autocorrelation (0.10) | Coefficient of variation (0.11) |
| 7 | Kurtosis (-0.04) | Skewness (0.06) | Skewness (0.10) | Skewness (0.10) |
| 8 | Skewness (-0.08) | Kurtosis (-0.02) | Kurtosis (0.02) | Kurtosis (0.03) |

With an imperfect diagnostic test, the correlation with emergence was more influenced by the noise structure (static vs. dynamical) than the noise magnitude (Table 4). For an imperfect test with 90% sensitivity and specificity, the correlation between all EWS metrics and emergence was relatively unaffected by the magnitude of static noise. The top four metrics with a perfect diagnostic test (autocovariance, variance, mean, and index of dispersion) maintained their positions as the most strongly correlated metrics.

For simulations with rubella-like SEIR dynamical noise, the correlation of all metrics was lower at low dynamical noise compared to low static noise (Table 4). With low levels of dynamical noise, the autocovariance, variance, and mean remained the most correlated with emergence ($|AUC - 0.5| = 0.16, 0.14$, and $0.13$, respectively). At high dynamical noise, these correlations disappeared, with all EWS metrics exhibiting $|AUC - 0.5| \leq 0.05$.

A full characterization of the strength of association between each metric and emergence, across all diagnostic tests and noise structures, can be seen in supplement (Supplemental Figure 2, Supplemental Figure 3, Supplemental Figure 4, Supplemental Figure 5).

Table 4: |AUC − 0.5| for EWS metrics, ranked in descending order of magnitude, computed on the subset of the emergent time series after the burn-in period, for a perfect test and an imperfect diagnostic test with 90% sensitivity and 90% specificity, under high and low static and dynamical noise systems

| Rank | Perfect Test | 90% Sensitive & Specific Imperfect Test | | | |
| | All Noise | Poisson Noise | | Dynamical Noise | |
| | | Low | High | Low | High |
| --- | --- | --- | --- | --- | --- |
| 1 | Autocovariance (0.20) | Autocovariance (0.23) | Autocovariance (0.22) | Autocovariance (0.16) | Mean (0.05) |
| 2 | Variance (0.20) | Variance (0.21) | Mean (0.20) | Variance (0.14) | Variance (0.04) |
| 3 | Mean (0.18) | Mean (0.20) | Variance (0.18) | Mean (0.13) | Autocovariance (0.03) |
| 4 | Index of dispersion (0.13) | Index of dispersion (0.17) | Index of dispersion (0.18) | Index of dispersion (0.09) | Coefficient of variation (0.02) |
| 5 | Autocorrelation (0.12) | Autocorrelation (0.17) | Coefficient of variation (0.17) | Autocorrelation (0.07) | Skewness (0.01) |
| 6 | Coefficient of variation (0.11) | Coefficient of variation (0.10) | Autocorrelation (0.16) | Skewness (0.06) | Autocorrelation (0.01) |
| 7 | Skewness (0.10) | Skewness (0.08) | Skewness (0.10) | Coefficient of variation (0.05) | Kurtosis (0.01) |
| 8 | Kurtosis (0.03) | Kurtosis (0.05) | Kurtosis (0.07) | Kurtosis (0.01) | Index of dispersion (0.00) |

## Predictive Ability

Each alert scenario (the combination of diagnostic test, noise structure and magnitude, and EWS metric) produced its optimal accuracy at a different combination of EWS hyperparameters (the quantile threshold of the long-running metric distribution to be exceeded to return a flag, and the number of consecutive flags required to trigger an alert) (Supplemental Figure 6, Supplemental Figure 7, Supplemental Figure 8, Supplemental Figure 9). At their respective maximal accuracies, the relative ranking of the EWS metrics computed with a perfect diagnostic test remained consistent to the ranking based upon |AUC − 0.5|: mean (accuracy = 0.72), variance (0.72), autocovariance (0.70), index of dispersion (0.63), autocorrelation (0.62), skewness (0.60), kurtosis (0.58), and coefficient of variation (0.50) (Supplemental Table 2).

When EWS metrics were computed on time series generated from imperfect diagnostic tests, each metric's accuracy generally remained constant, with a few notable exceptions (Figure 1, Supplemental Figure 1). For the 4 most correlated metrics (autocovariance, variance, mean, and index of dispersion), the accuracy achieved with imperfect diagnostic tests was comparable for low and high static noise, for all diagnostic test accuracies (Figure 1). The accuracy of outbreak detection using index of dispersion increased with decreasing diagnostic test sensitivity and specificity for low and high levels of static noise (Figure 1, Supplemental Figure 6, Supplemental Figure 7). For low dynamical noise, accuracy increased slightly for diagnostic test sensitivity and specificity greater than 97% and then declined (Figure 1). For high dynamical noise, accuracy declined monotonically with decreasing test sensitivity and specificity (Figure 1, Supplemental Figure 9). Results for the 4 least well correlated EWS metrics are presented in the supplement (Supplemental Figure 1).

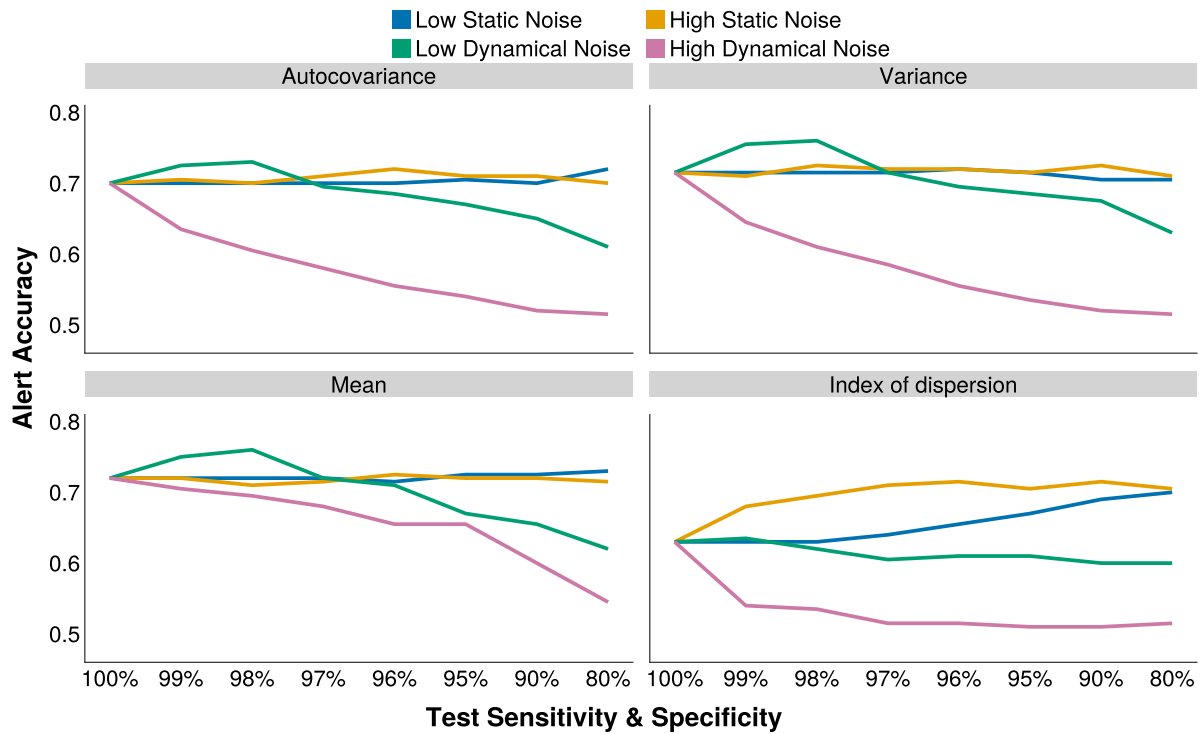

Figure 1: The change in alert accuracy for the most correlated EWS metrics under increasing diagnostic uncertainty, and low and high levels of static or dynamical noise. Low noise refers to simulations where the average incidence of noise is equal to the average incidence of measles. High noise refers to simulations where the average incidence of noise is equal to 7 times the average incidence of measles. The tests sensitivity equals the test specificity for all diagnostic tests.

Outbreak detection produced false positives under the null simulations for all EWS metrics, except for the coefficient of variation computed on time series resulting from perfect tests, which also failed to alert in emergent simulations. Here we illustrate the comparison of timing of alerts for the autocovariance metric for the null and emergent simulations (Figure 2). Outbreak detection using the autocovariance metric resulted in comparable timing of alerts for perfect and imperfect tests under low and high static noise (Figure 2). For low dynamical noise, the imperfect test resulted in a similar number of true positives under the emergent scenario, but tended to trigger those alerts later than with a perfect test. Notably, an imperfect test resulted in more false positives under the null scenario and tended to trigger those alerts later. With high dynamical noise, an imperfect test failed to produce many alerts under either the null or emergent scenarios (Figure 2).

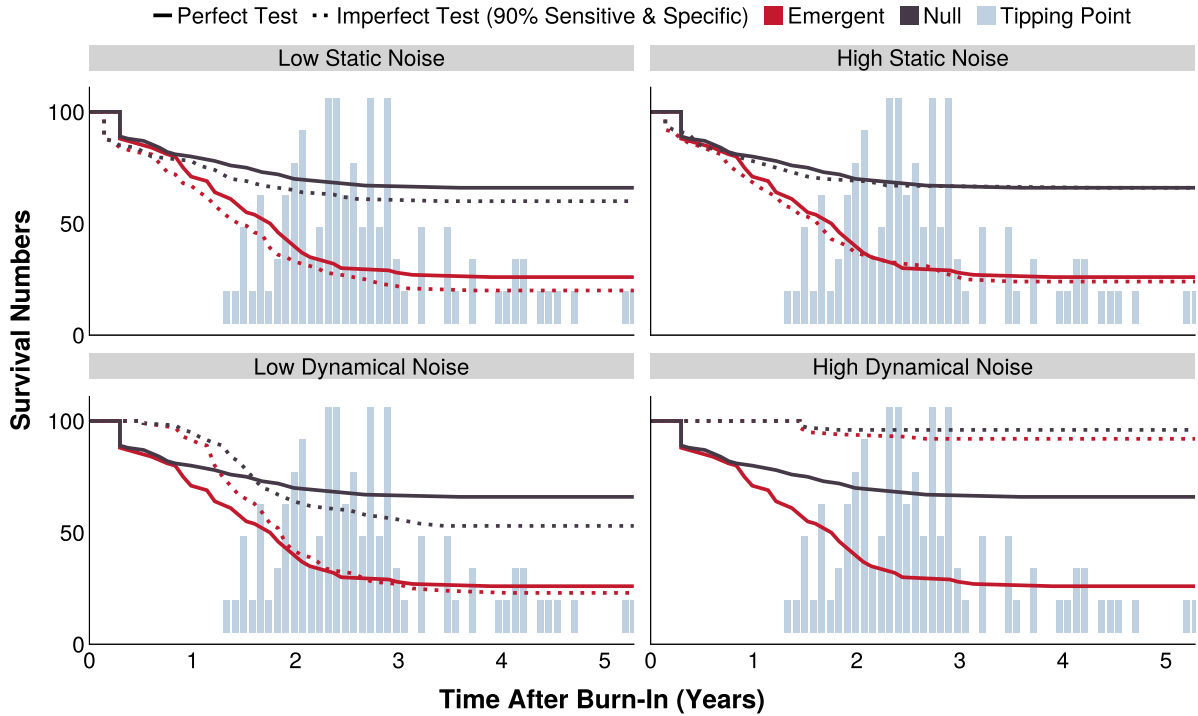

Figure 2: Survival curves for the autocovariance EWS metric computed on emergent and null simulations, with a perfect test, and an imperfect test that is 90% sensitive and specific. The histogram depicts the times when the tipping point is reached ($R_{\mathrm{E}} = 1$) under the emergent simulation, right-truncating the curves. The trajectory of the solid lines are identical in each facet, as the perfect test is unaffected by noise cases. The histogram is identical between facets as it represents the timing of the tipping points, and all testing scenarios use the same underling measles simulations that terminate at the tipping point

## Discussion

Outbreak detection using EWS metrics is robust to diagnostic uncertainty depending on the structure and magnitude of the noise due to non-target infections in the surveillance time series. Under static noise, outbreak detection using a time series of test positive cases resulted in similar performance to a corresponding time series with a perfect diagnostic, regardless of the incidence of the non-target infections. However, when the background noise due to non-target infections in the time series of suspected cases is consistent with a dynamical SEIR-type process (e.g., tends to produce cycles or periods of consistent outbreaks), the accuracy of outbreak detection declines with decreasing diagnostic test sensitivity and specificity and with increasing relative incidence of the non-target infections. Thus, the performance of outbreak warning systems using EWS depends on both the properties of the individual diagnostics used and structure and magnitude of non-target disease incidence, which may vary with the local context.

This analysis was motivated by the case of anticipating the progression to $R_{\mathrm{E}} > 1$ for measles in the context of other sources of febrile rash; e.g., rubella, parvovirus, and arboviruses such as dengue fever and chikungunya [44,50–52]. For much of the WHO's African Region, the co-circulation of measles and rubella is common, although there are stark differences in the relative proportion of incidence by country [36]. In Guinea Bissau, for example, the estimated incidence rates of rubella are approximately 9 time that of measles, in Botswana they are similar, and in the Democratic Republic of Congo measles incidence is estimated to be 20 times higher [36]. Imperfect diagnostic tests will not provide equal value to each of these locations. The Democratic Republic of Congo would be a good candidate for the integration of less accurate diagnostic if it allowed for improvements to other aspects of the disease surveillance system e.g., increases in the testing rates and case-finding activities due to lower financial and logistical costs [53]. However, when large rubella outbreaks can produce meaningful peaks in test positive cases resulting from the use of imperfect diagnostics, such as in Guinea Bissau, the EWS metrics struggle to discriminate between emergent and null periods, reducing their utility.

When evaluating the ability for the EWS metrics to accurately discriminate between emergent and null simulations, it is import to contextualize the results with the system's relative speed and specificity. Alert systems necessarily make compromises in their design: improvements to speed generally come at the cost of increased numbers of false alerts [54,55]. Depending on the context, it may be desirable to place a greater weight in preference/penalty for one of these axes; in scenarios where the expected cost to launch a preliminary investigation is low relative to the unaverted DALYs resulting from incorrect inaction in an overly specific system, higher false alert rates may be acceptable. This analysis provides a framework to explicitly explore these trade-offs, including through the comparison of survival curves. A larger separation at the end of the time series between the emergent and null simulation lines indicates higher alert accuracy, as there is a greater difference in the true positive and false positive rates. Faster and more sustained declines indicate a (relatively) more sensitive alert system with more advanced warning of emergence.

Under the simulation constraints placed here, generally, the use of imperfect diagnostic tests does not increase the speed of the warning for the EWS metrics that are predictive of emergence. This is likely a consequence of imperfect diagnostic tests producing more false positive cases, which, without appropriate penalization, would otherwise lead to high false alert rates under the same EWS hyperparameters. Adjusting the relative weighting of alert sensitivity and specificity used to compute the alert accuracy would allow for an exploration of alternative scenarios. Additionally, multiple hyperparameter combinations can produce identical alert accuracy i.e., some combinations will favor the speed of alert at the expensive of its specificity, and *vice versa*. In situations where identical accuracies are achieved, we present the results associated with the most specific alert system i.e., reduce the number of false positive alerts. An equivalent set of analyses could be computed to favor the hyperparameter combinations that produced the fastest and most frequent alerts.

For EWS metrics to reflect the underlying dynamics of critical slowing down, careful detrending of the data is required [21,56,57]. Our analysis utilizes a backward-facing uniform moving average to detrend the data: this was chosen as it can be easily intuited and implemented in a surveillance system. However, it has previously been stated that insufficient detrending may lead to spurious patterns that do not arise from a system's dynamical response [21], with some EWS more sensitive than others, and that the associations may vary with the bandwidth size selected [56,57]. While it may be preferred to detrend using the mean over multiple realizations [56], this is clearly not possible in a real-world situation. Future work could explore the effects of different detrending methods (e.g., Gaussian weighting moving average, smaller and/or larger bandwidths) on the effectiveness of EWS metrics in systems with diagnostic uncertainty [56,57]. Similarly, prior work has demonstrated the benefits of constructing composite metrics, for example, calculating a composite metric as the sum of the standardized differences for each of the individual metrics, before defining a threshold for the distribution of the composite [22]. However, there are other techniques that could be applied to produce a composite statistic [29,30], each requiring different decisions as to the appropriate weightings to be assigned to the underlying single metrics. For this reason, we only present the results from the individual metrics to illustrate the effects of diagnostic uncertainty, but future work should aim to extend the approach detailed to composite EWS metrics.

Despite being relatively well-established in areas of study such as ecology, ecosystem collapse, and climate science [19,21–23,32,58,59], the exploration and development of EWS for infectious disease systems is in its relative infancy. Until recently, a large proportion of the prior work in the area has been to establish the existence of these metrics that theoretically could be used in such a system [15,25,26]. While this is a crucial first step, for use in a proactive outbreak alert system, EWS metrics must be able to provide advance warning of the approach to the tipping point $R_{\mathrm{E}} = 1$. Correlations alone are not sufficient to indicate when and what actions must be taken. To address this, there is a growing body of work that seeks to evaluate the use of various threshold and risk-based approaches within infectious disease systems [27,29,30,35]. Our work expands upon these efforts, characterizing the limits of predictability for EWS metrics in systems with diagnostic uncertainty and background noise.

# Funding

This work was supported by the Bill & Melinda Gates Foundation through grant No. INV-016091_2020. This project was also supported by the National Science Foundation, by grant: NSF-NIH-NIFA Ecology and Evolution of Infectious Disease award DEB 1911962. The funding sources had no role in the collection, analysis, interpretation, or writing of the work.

# Acknowledgements

## Author Contributions

*Conceptualization:* CA, MJF

*Data curation:* CA, MJF

*Formal analysis:* CA

*Funding acquisition:* MJF

*Investigation:* CA

*Methodology:* CA, MJF

*Writing - original draft:* CA

*Writing - review and editing:* all authors.

## Conflicts of Interest and Financial Disclosures

## Data Access, Responsibility, and Analysis

## Data Availability

All code and data for the simulations can be found at https://github.com/arnold-c/CSDNoise

# Diagnostic Uncertainty Limits the Potential of Early Warning Signals for Epidemic Transitions

Callum R.K. Arnold[1,2,*], Matthew J. Ferrari[1,2]

[1] Department of Biology, Pennsylvania State University, University Park, PA, USA 16802
[2] Center for Infectious Disease Dynamics, Pennsylvania State University, University Park, PA, USA 16802

* Corresponding author. Callum R.K. Arnold. Address: Department of Biology, Pennsylvania State University, University Park, PA, USA 16802. Email: contact@callumarnold.com.

# Figures

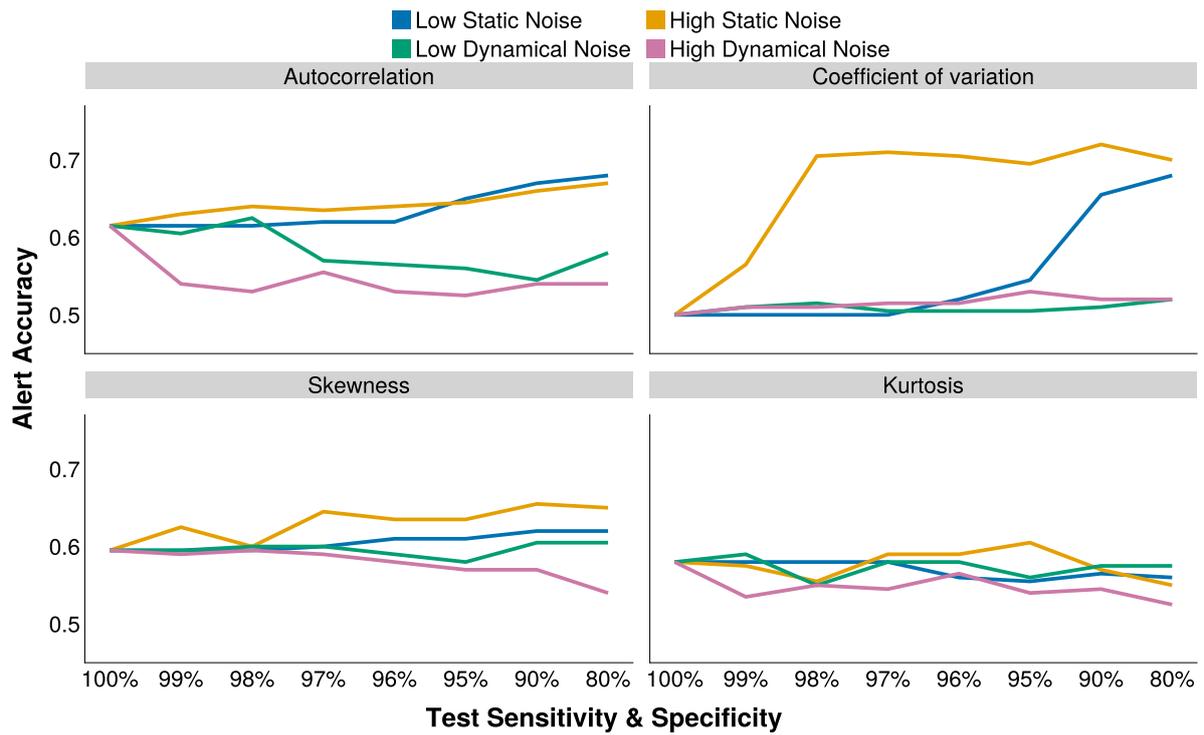

Supplemental Figure 1: The change in alert accuracy for the least correlated EWS metrics under increasing diagnostic uncertainty, and low and high levels of static or dynamical noise. Low noise refers to simulations where the average incidence of noise is equal to the average incidence of measles. High noise refers to simulations where the average incidence of noise is equal to 7 times the average incidence of measles. The test sensitivity equals the test specificity for all diagnostic tests.

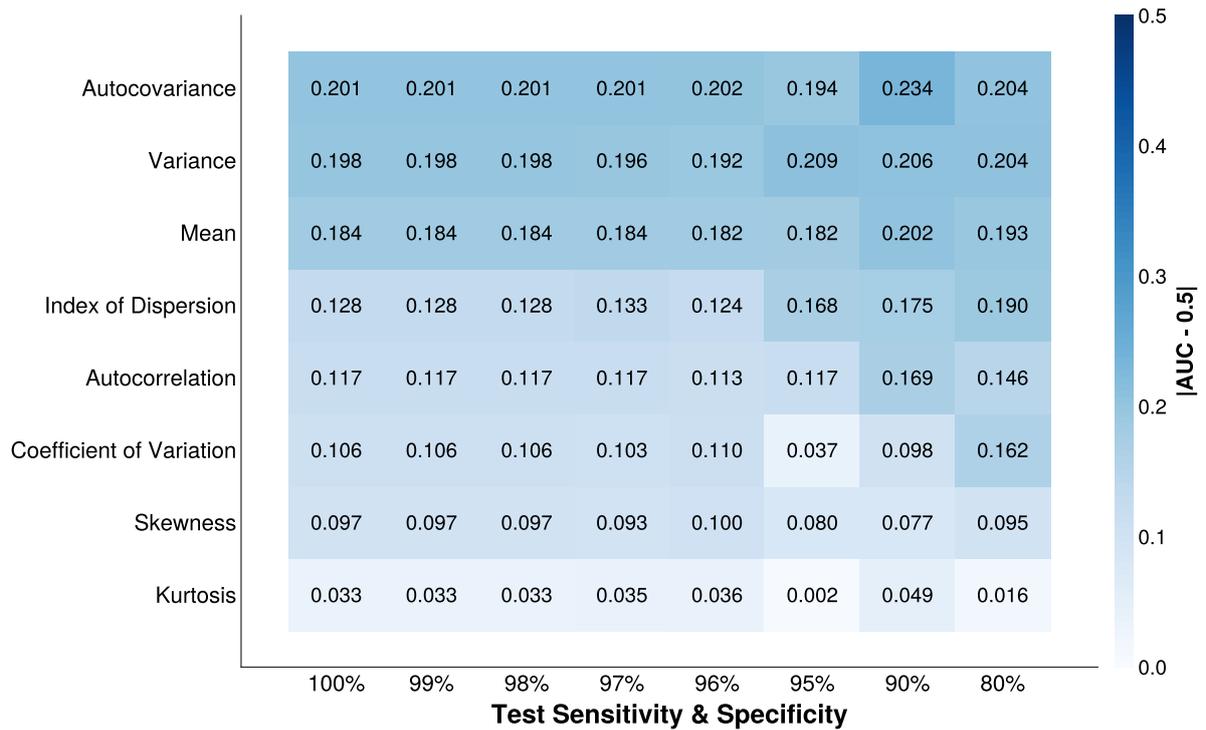

Supplemental Figure 2: The strength of the correlation ($|AUC - 0.5|$) for each EWS metric with emergence, at low levels of static noise, for diagnostic tests of varying accuracy, and was computed after the completion of the burn-in period. The test sensitivity equals the test specificity for all diagnostic tests.

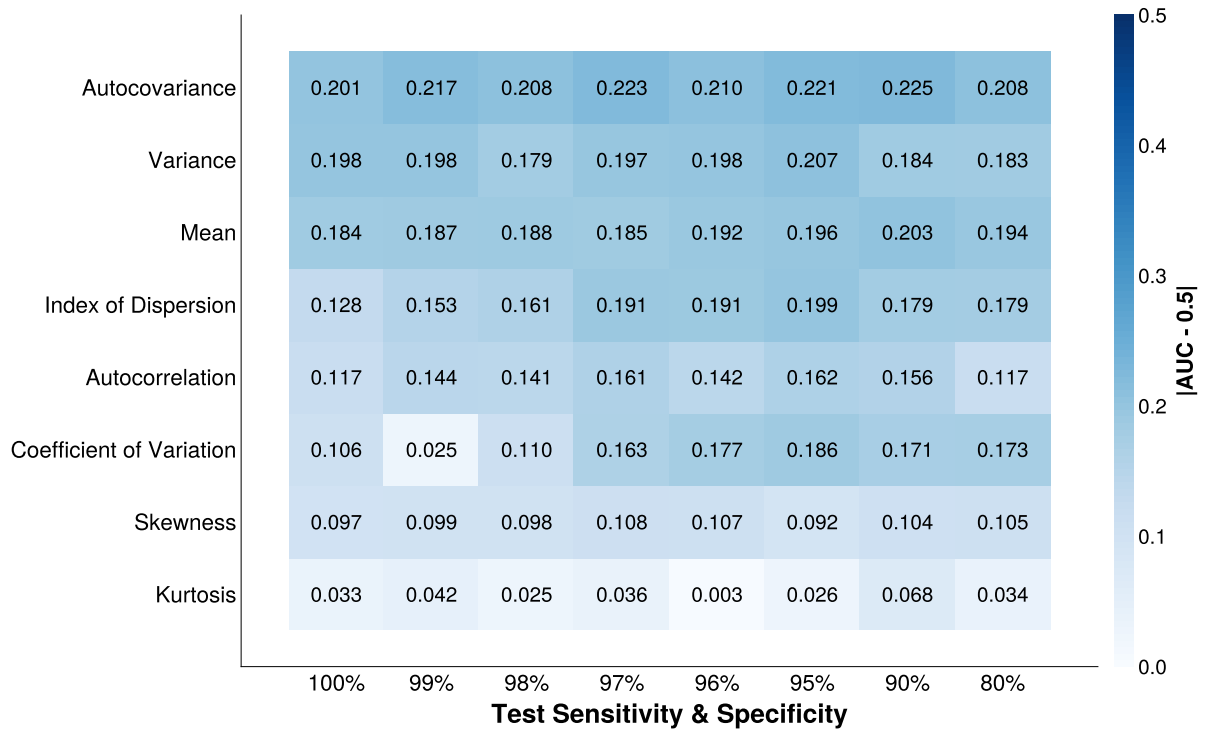

Supplemental Figure 3: The strength of the correlation ($|AUC - 0.5|$) for each EWS metric with emergence, at high levels of static noise, for diagnostic tests of varying accuracy, and was computed after the completion of the burn-in period. The test sensitivity equals the test specificity for all diagnostic tests.

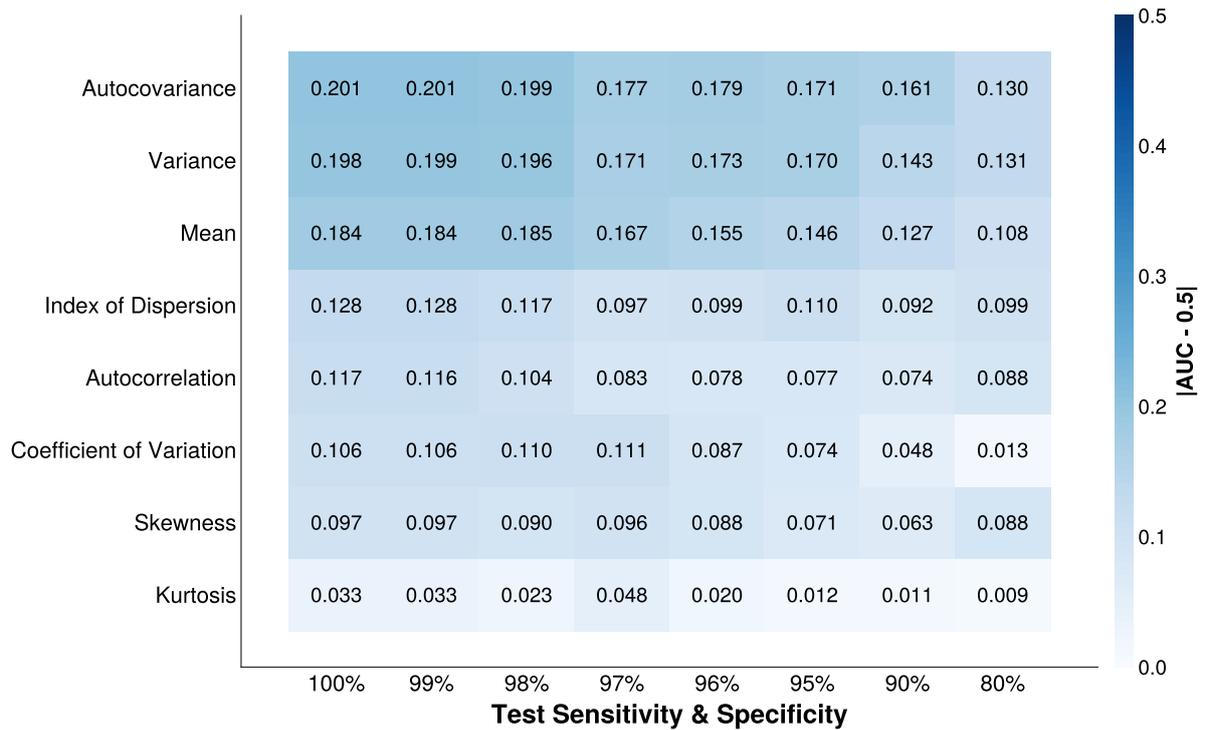

Supplemental Figure 4: The strength of the correlation ($|\text{AUC} - 0.5|$) for each EWS metric with emergence, at low levels of dynamical noise, for diagnostic tests of varying accuracy, and was computed after the completion of the burn-in period. The test sensitivity equals the test specificity for all diagnostic tests.

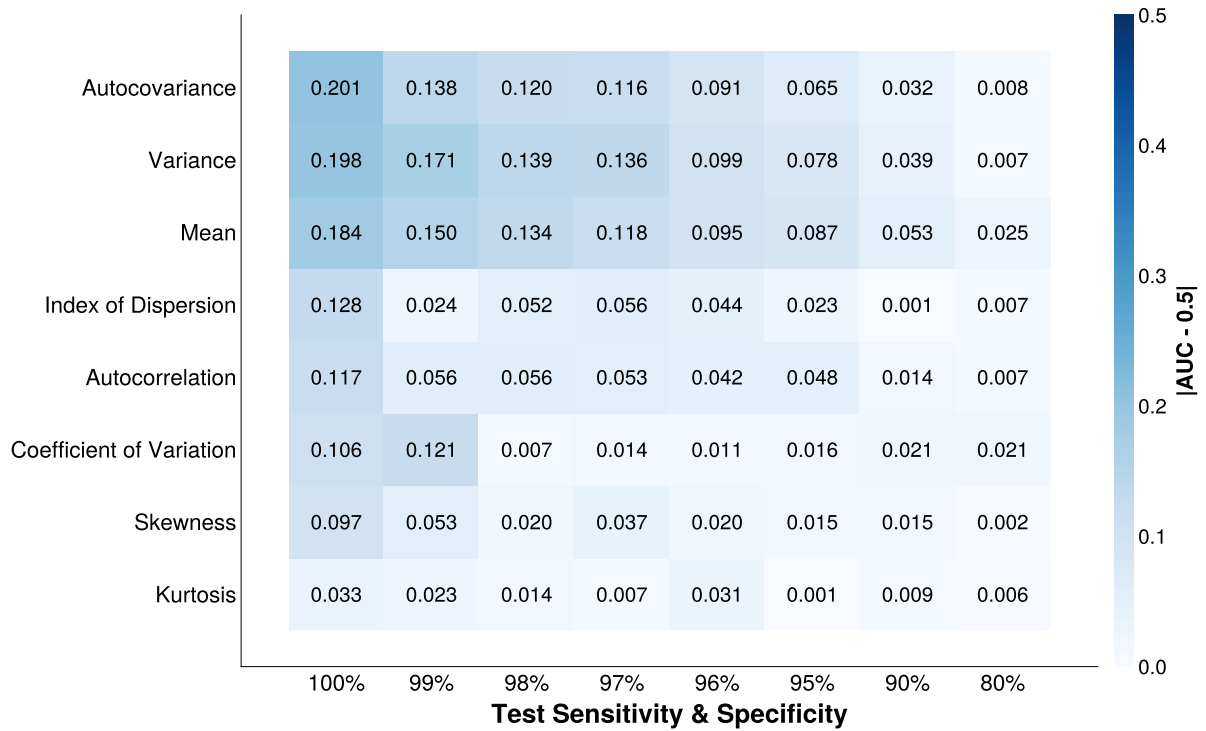

Supplemental Figure 5: The strength of the correlation ($|\text{AUC} - 0.5|$) for each EWS metric with emergence, at high levels of dynamical noise, for diagnostic tests of varying accuracy, and was computed after the completion of the burn-in period. The test sensitivity equals the test specificity for all diagnostic tests.

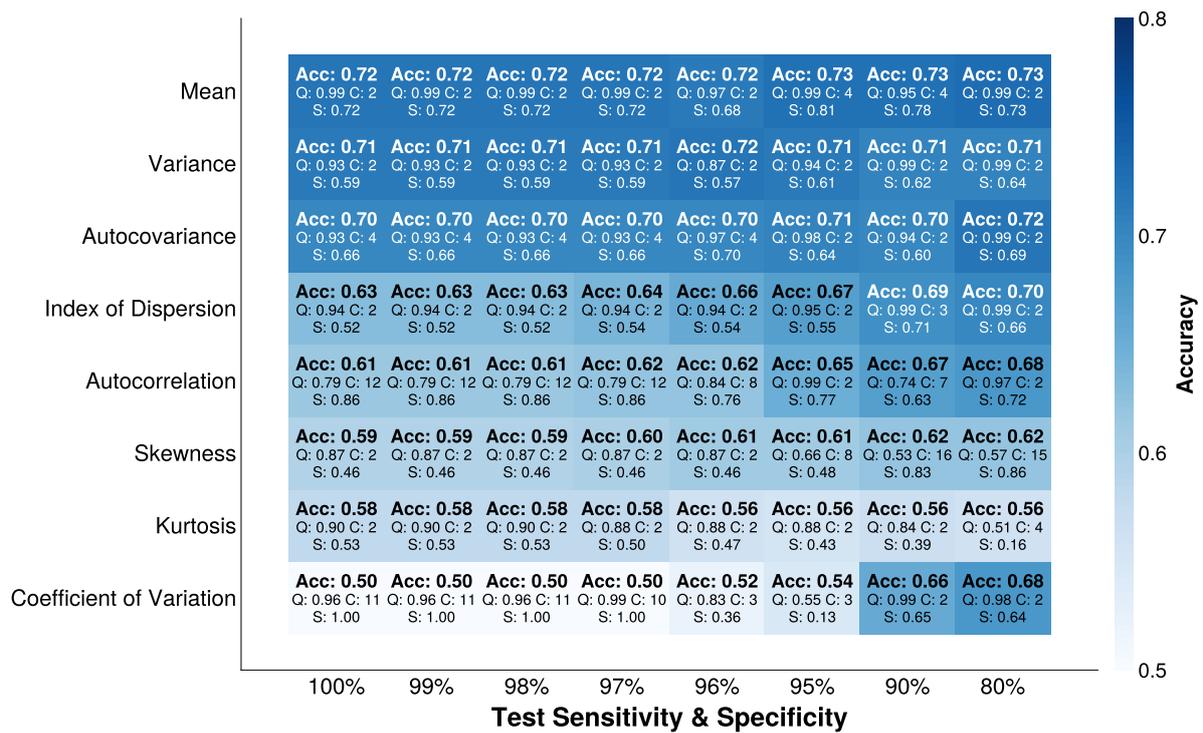

Supplemental Figure 6: The maximal alert accuracy achieved by each EWS metric under low levels of static noise. Q) refers to the long-running quantile threshold to return a flag, and C) the number of consecutive flags to trigger an alert, that in combination produce the maximal accuracy. S) refers to the resulting specificity of the alert system. The test sensitivity equals the test specificity for all diagnostic tests.

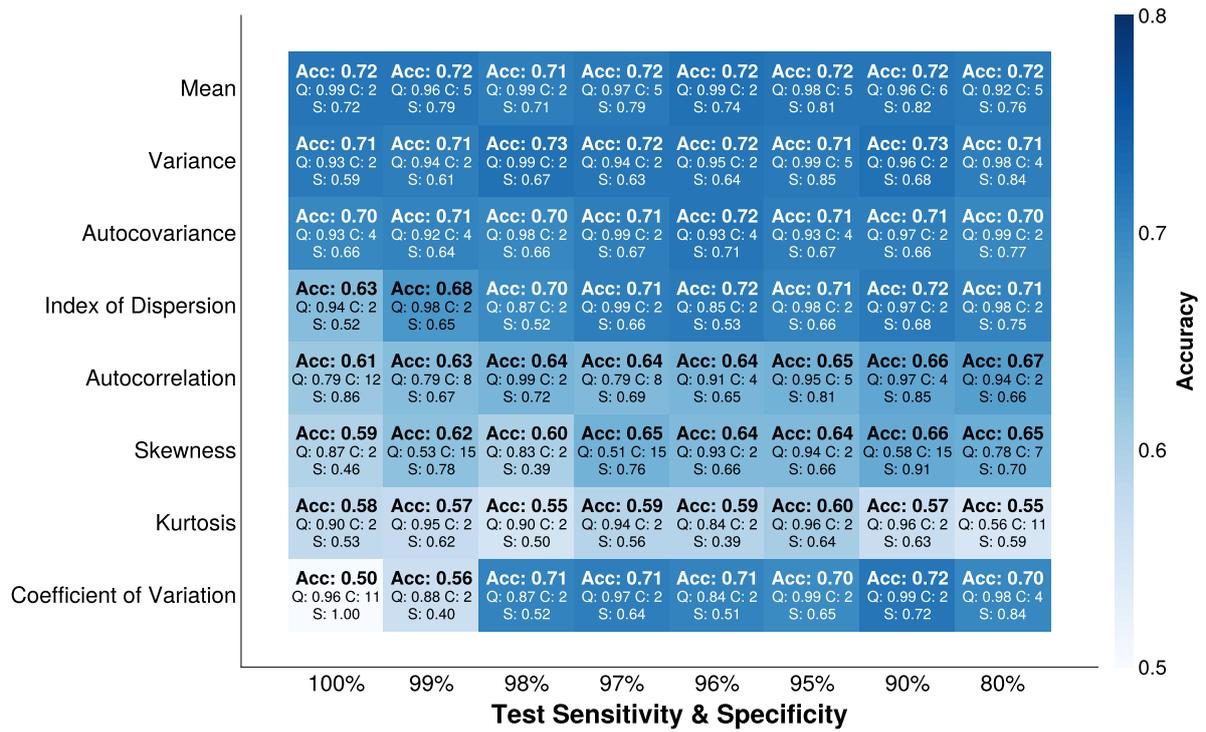

Supplemental Figure 7: The maximal alert accuracy achieved by each EWS metric under high levels of static noise. Q) refers to the long-running quantile threshold to return a flag, and C) the number of consecutive flags to trigger an alert, that in combination produce the maximal accuracy. S) refers to the resulting specificity of the alert system. The test sensitivity equals the test specificity for all diagnostic tests.

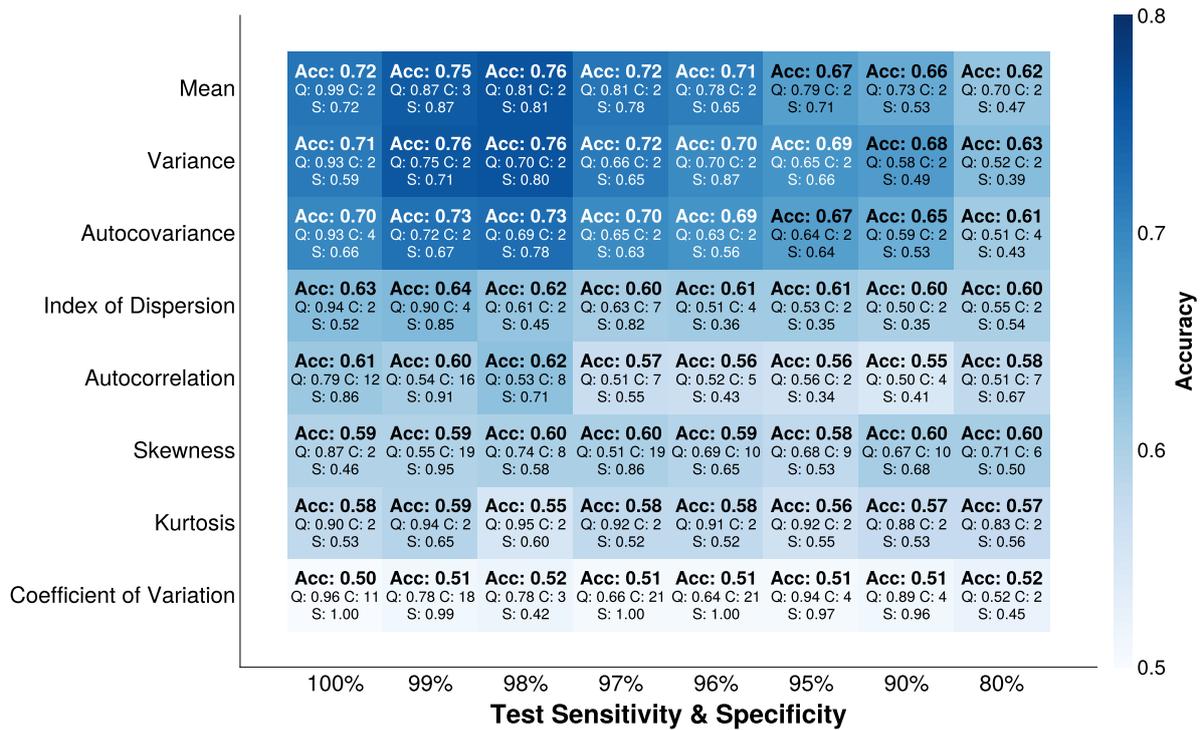

Supplemental Figure 8: The maximal alert accuracy achieved by each EWS metric under low levels of dynamical noise. Q) refers to the long-running quantile threshold to return a flag, and C) the number of consecutive flags to trigger an alert, that in combination produce the maximal accuracy. S) refers to the resulting specificity of the alert system. The test sensitivity equals the test specificity for all diagnostic tests.

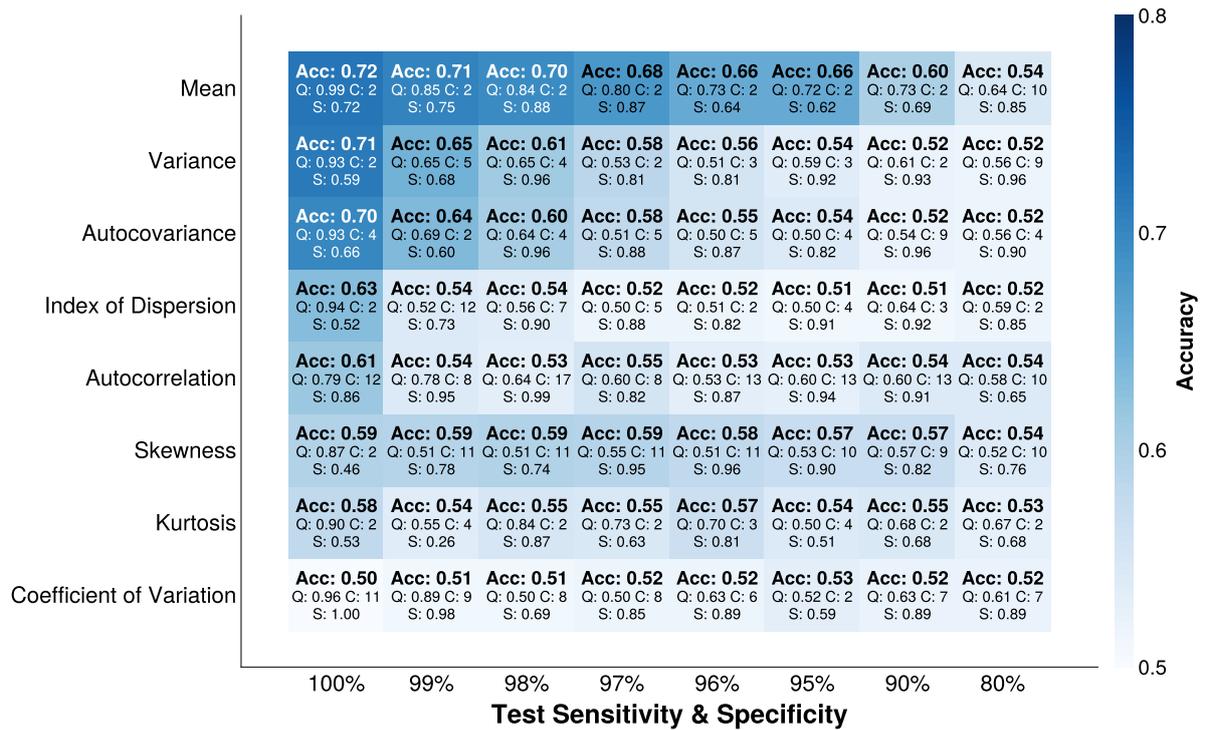

Supplemental Figure 9: The maximal alert accuracy achieved by each EWS metric under high levels of dynamical noise. Q) refers to the long-running quantile threshold to return a flag, and C) the number of consecutive flags to trigger an alert, that in combination produce the maximal accuracy. S) refers to the resulting specificity of the alert system. The test sensitivity equals the test specificity for all diagnostic tests.

## Tables

Supplemental Table 1: The ranking and mean value of Kendall's Tau computed on the subset of the emergent time series after the burn-in period, for a perfect test and an imperfect test with sensitivity and specificity equal to 90%, under high and low static and dynamical noise systems

| Rank | Perfect Test | 90% Sensitive & Specific Imperfect Test | | | |
| --- | --- | --- | --- | --- | --- |
| | All Noise | Static Noise | | Dynamical Noise | |
| | | Low | High | Low | High |
| 1 | Variance (0.62) | Variance (0.61) | Variance (0.60) | Variance (0.66) | Index of dispersion (0.47) |
| 2 | Index of dispersion (0.58) | Index of dispersion (0.59) | Index of dispersion (0.60) | Autocovariance (0.63) | Autocorrelation (0.45) |
| 3 | Autocovariance (0.58) | Autocovariance (0.55) | Coefficient of variation (0.59) | Index of dispersion (0.57) | Coefficient of variation (0.45) |
| 4 | Autocorrelation (0.38) | Coefficient of variation (0.51) | Autocovariance (0.51) | Mean (0.48) | Autocovariance (0.39) |
| 5 | Mean (0.38) | Autocorrelation (0.41) | Mean (0.37) | Autocorrelation (0.42) | Variance (0.38) |
| 6 | Coefficient of variation (0.15) | Mean (0.35) | Autocorrelation (0.36) | Coefficient of variation (0.12) | Skewness (0.11) |
| 7 | Skewness (0.06) | Skewness (0.14) | Skewness (0.10) | Skewness (-0.05) | Kurtosis (-0.19) |
| 8 | Kurtosis (-0.02) | Kurtosis (0.01) | Kurtosis (0.02) | Kurtosis (-0.11) | Mean (-0.21) |

Supplemental Table 2: The ranking and alert accuracy of the EWS-based alert system computed on the subset of the emergent time series after the burn-in period, for a perfect test and an imperfect test with sensitivity and specificity equal to 90%, under high and low static and dynamical noise systems

| Rank | Perfect Test | 90% Sensitive & Specific Imperfect Test | | | |
| --- | --- | --- | --- | --- | --- |
| | All Noise | Static Noise | | Dynamical Noise | |
| | | Low | High | Low | High |
| 1 | Mean (0.72) | Mean (0.73) | Variance (0.73) | Variance (0.68) | Mean (0.60) |
| 2 | Variance (0.71) | Variance (0.71) | Coefficient of variation (0.72) | Mean (0.66) | Skewness (0.57) |
| 3 | Autocovariance (0.70) | Autocovariance (0.70) | Mean (0.72) | Autocovariance (0.65) | Kurtosis (0.55) |
| 4 | Index of dispersion (0.63) | Index of dispersion (0.69) | Index of dispersion (0.72) | Skewness (0.60) | Autocorrelation (0.54) |
| 5 | Autocorrelation (0.61) | Autocorrelation (0.67) | Autocovariance (0.71) | Index of dispersion (0.60) | Autocovariance (0.52) |
| 6 | Skewness (0.59) | Coefficient of variation (0.66) | Autocorrelation (0.66) | Kurtosis (0.57) | Coefficient of variation (0.52) |
| 7 | Kurtosis (0.58) | Skewness (0.62) | Skewness (0.66) | Autocorrelation (0.55) | Variance (0.52) |
| 8 | Coefficient of variation (0.50) | Kurtosis (0.56) | Kurtosis (0.57) | Coefficient of variation (0.51) | Index of dispersion (0.51) |